\documentclass[twoside]{article}
\usepackage{fleqn,espcrc2,epsfig}

\newcommand{\AmS}{{\protect\the\textfont2  
   A\kern-.1667em\lower.5ex\hbox{M}\kern-.125emS}}  
   
\newcommand{\Dslash}{{\not{\hspace{-0.10cm}D}}}
\newcommand{\pslash}{{\not{\hspace{-0.07cm}p}}}

\newcommand{\half}{\frac{1}{2}}
\newcommand{\calO}{ {\cal O} }

\def\3{\ss}

\title{
\vspace{-2.2cm}
\flushleft{\normalsize TPR 99-18} \hfill\\
\vspace{-0.5cm}
\flushleft{\normalsize HUB-EP-99/52} \hfill\\
\vspace{-0.4cm}
\flushleft{\normalsize September 1999} \hfill\\
\vspace{0.7cm}
On-shell and off-shell improvement for Ginsparg-Wilson fermions%
 \thanks{Talk given by P.E.L. Rakow at  Lattice '99, Pisa, Italy.} }

\author{ S.~Capitani
 \address{Deutsches Elektronen-Synchrotron DESY,
 D-22603 Hamburg, Germany}%
 , M.~G\"ockeler \address{
  Institut f\"ur Theoretische Physik, Universit\"at Regensburg,
 D-93040 Regensburg, Germany}%
 , R.~Horsley \address{Institut f\"ur Physik, Humboldt-Universit\"at zu Berlin,
 D-10115 Berlin, Germany}%
 , P.E.L.~Rakow$\,^{\rm b}$ 
 and G.~Schierholz$\,\,^{\rm a,}$\address{  
 Deutsches Elektronen-Synchrotron DESY 
 and NIC, D-15735 Zeuthen, Germany }}
    
\begin{document}

 \begin{abstract}  
 We discuss the improvement of bilinear fermionic operators for
 Ginsparg-Wilson fermions. 
  We present explicit formulae for improved Green's functions,
 which apply both on-shell and off-shell.  
 \end{abstract}  
   
\maketitle


\section{ Introduction} 

  A fermion action that fulfills the Ginsparg-Wilson condition
 realises chiral symmetry in an unconventional way that allows it
 to avoid the Nielsen-Ninomiya theorem \cite{GW,Luescher}. This
 chiral symmetry  automatically ensures that the hadron masses of
 the theory are free of  $O(a)$ discretisation errors.

  However, if we are interested in going beyond
 this and calculating matrix elements (for example structure functions, 
 decay constants etc.) we also need to know how to improve fermion 
 operators. When we measure hadronic matrix elements, it is enough if the 
 operators are improved for on-shell quantities. 
 To do non-perturbative renormalisation we may want to
 measure off-shell Green's functions with a virtuality large enough that
 we can reasonably compare with continuum perturbation theory, so 
 improvement of off-shell Green's functions is useful too. 
   
 \section{ Ginsparg-Wilson fermions }

 The continuum Dirac operator, $\Dslash$, anticommutes with $\gamma_5$, 
 but we know from the Nielsen-Ninomiya theorem that a lattice
 Dirac operator cannot do this without having problems such as 
 doubling or non-locality. Ginsparg-Wilson fermions~\cite{GW}
 are defined by the anti-commutation relation
 \begin{equation} 
    D_{GW} \, \gamma_5 + \gamma_5 \, D_{GW}
  = a  D_{GW} \,  \gamma_5  \, D_{GW}. 
 \end{equation} 

    The standard way of writing down massive Ginsparg-Wilson
 fermions is to use the  fermion matrix~\cite{NiederReview}  
 \begin{equation} 
  M \equiv \left(1 - \frac{1}{2} a m_0 \right) D_{GW} + m_0 .
 \end{equation} 
 $M$ is local and always invertible. 

 Following~\cite{Zenkin}, we can define the associated matrix 
 \begin{equation} 
  K_{GW}  \equiv  
 \left( 1 - \frac{a}{2} D_{GW}\right)^{-1} D_{GW}.
 \label{Kdef}
 \end{equation} 
 The Ginsparg-Wilson condition implies that 
 \begin{equation} 
  K_{GW} \, \gamma_5 + \gamma_5 \, K_{GW} = 0 . 
 \end{equation} 
 Thus $K_{GW}$ has the same chiral properties as the continuum
 Dirac operator. Also, the eigenvalues of $K_{GW}$ are 
 imaginary, like the eigenvalues of the continuum $\Dslash$. 
 The fermion propagator we really want to find 
 is the propagator calculated with $K_{GW}$.

 \section{ Fermion propagator } 

 The unimproved fermion propagator is obtained simply 
 by inverting $M$. 
  \begin{equation} 
  S \equiv \langle \psi \bar{\psi} \rangle 
 = \frac{1}{a^4} \left\langle M^{-1} \right\rangle  . 
 \end{equation} 
 However this propagator has unwanted contact terms which are of $O(a)$. 

 We would prefer the improved ($\star$) propagator 
 defined from the matrix $K_{GW}$~\cite{Chiu}:  
 \begin{equation} 
  S_\star = \frac{1}{a^4} \left\langle \frac{1}{K_{GW} +m_0}\right\rangle .
 \end{equation} 
 Since $K_{GW}$ has the same chiral properties as the continuum 
 Dirac operator, the improved propagator $S_\star$ is automatically 
 free from all $O(a)$ discretisation errors.  However, we need a
 way to calculate $S_\star$ directly from $M$, without having to 
 construct the problematic $K_{GW}$. Substitute eq.~(\ref{Kdef}), 
 the definition of $K_{GW}$, into the above equation,
 and a little algebra gives: 
 \begin{equation} 
   S_\star (x,y) = \frac{1}{1 + a m_0 b_\psi }
 \left( S(x,y) - \frac{a}{2} \lambda_\psi \delta (x-y) \right) 
 \label{simp} 
 \end{equation} 
 with improvement coefficients
 \begin{equation}
 b_\psi =  - \, \frac{1}{2}
 \hspace{0.5cm}{\rm and} \hspace{0.5cm}
 \lambda_\psi = 1 .
 \label{propimpco}
 \end{equation}
 Note that we never have to explicitly construct $K_{GW}$, the 
 only matrix we invert is $M$, so the  
 formula is applicable even when $D_{GW}$ has eigenvalues at $2 /a$. 
  The chiral violation in $S$ is concentrated in a $\delta$ 
 function contact term at the origin, so it only gives a problem 
 when we are interested in off-shell quantities.  

 To illustrate the difference between the unimproved and improved
 propagators we will show how improvement works in the case of
 the free fermion theory.  
  Starting from the massless Wilson fermion matrix $D_W$,
 Neuberger~\cite{Neuberger} introduces the matrix $A$, defined by
 \begin{equation} 
  A \equiv 1 - a D_W . 
 \end{equation} 
 It can then be shown that the operator
 \begin{equation}
  D_N \equiv \frac{1}{a} 
 \left( 1 - A / \sqrt{ A^\dagger A} \, \right)   
 \end{equation} 
 satisfies the Ginsparg-Wilson condition.

\begin{figure}[t]
\begin{center}
 \epsfig{file = 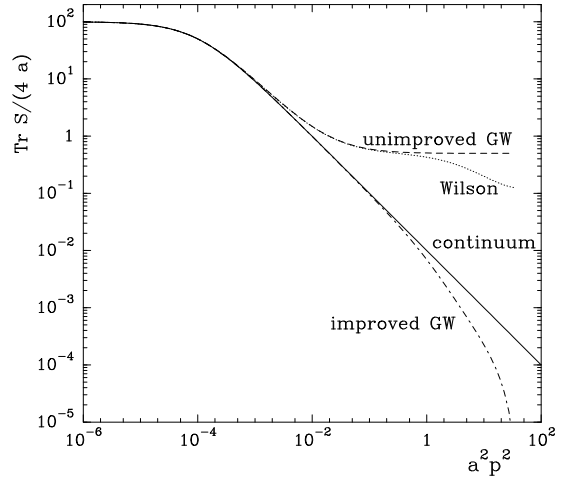, angle = 270, width = 7.2cm}
\end{center}
 \caption{The trace of the free fermion propagator for $a m_0 = 0.01$.
 Momentum is taken in the direction $(1,1,1,1)$.} 
 \label{propfig} 
 \end{figure}

 Although $D_N$ satisfies the Ginsparg-Wilson condition exactly, 
 it is not yet off-shell improved.  If we expand for small 
 momenta we find~\cite{GWletter} 
 \begin{equation} 
  D_N(p) = {\rm i} \pslash + \half a p^2 + O(a^2 p^3), 
 \end{equation} 
 which is no improvement over the Wilson propagator. 
 However, if we calculate the improved propagator of
 eq.~(\ref{simp}) we find that it has errors of $O(a^2)$. 
 This is illustrated in Fig.~\ref{propfig}, where we compare 
 the trace of the Wilson action propagator, the unimproved
 Ginsparg-Wilson propagator and the improved propagator 
 with the continuum result.  

 \section{Improving bilinear operators} 

 In this section we present the main conclusions from~\cite{GWletter}.  
 Let us consider improving the Green's function for
 a flavour non-singlet operator of the form 
 $\bar{\psi} \Gamma \tau D_\alpha \cdots D_\omega \psi$, where 
 $\Gamma$ is a matrix in the Clifford algebra,  
 $\tau$ a flavour matrix and $D_\alpha$ a covariant 
 derivative. 

 As in the case of the propagator, we know that the
 Green's function $G_\star^\calO$ 
  \begin{equation} 
  G_\star^\calO =  \frac{1}{a^4} \left\langle
   \frac{1}{K_{GW} + m_0 } O  \frac{1}{K_{GW} + m_0 }
 \right\rangle   
 \end{equation} 
 is free of $O(a)$ effects, but we want to express it
 in terms of the well-behaved matrix $M$. 
 One possibility is
 \begin{equation} 
  G_\star^\calO = \frac{1}{a^4} \; \; 
  \frac{1} { 1 + a m_0 b_\psi } \, 
 \left\langle M^{-1} \widetilde{O}  M^{-1} \right\rangle 
 \end{equation} 
 where 
 \begin{equation}
  \widetilde{O} = ( 1 + a m_0 b_\psi )
 (1 - \frac{a}{2} D_{GW} ) O (1 - \frac{a}{2} D_{GW} )
 \end{equation} 

  \begin{figure}[t]
\begin{center}
\epsfig{file = 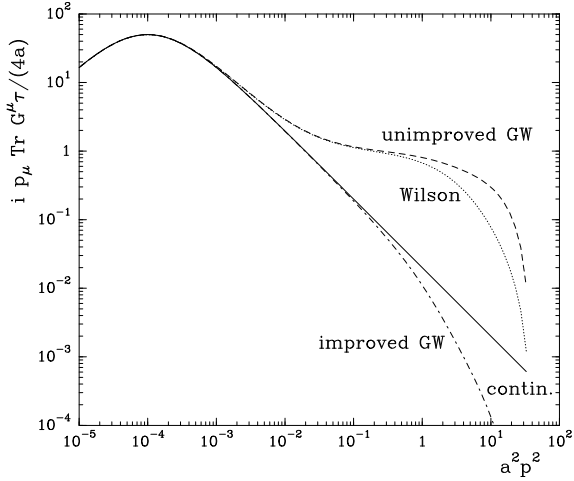, angle = 270, width = 7.5cm}
\end{center}
 \caption{
 Green's function improvement for the local vector
 current $\bar{\psi} \gamma_\mu \tau \psi$.}
 \label{fig3pt} 
 \end{figure} 

  This is not the most general expression for
 $G_\star^\calO$. We can use the identity
 \begin{equation} 
   D_{GW} M^{-1} = \frac{1}{1 - a m_0/2} \; 
 \left( - m_0 M^{-1} + \delta_{x y} \right) 
\end{equation} 
 to get the alternative expression 
 \begin{equation} 
  G_\star^\calO = \frac{1} { 1 + a m_0 b_\psi }
 \left[ G_\circ - \frac{a}{2} \lambda_\calO C^\calO \!
 + \frac{a^2}{4} \frac{\eta_\calO}{a^4} \langle O \rangle \right]  
 \label{gwimpsimp} 
\end{equation} 
  where
 \begin{eqnarray}
  G_\circ &\equiv & \frac{1}{a^4} \left\langle
 M^{-1} O_\star { M}^{-1} \right\rangle \nonumber \\
 O_\star &\equiv & O + a c_0  m_0 O
 - \frac{a}{2} c_1 ( D_{GW} O + O \, D_{GW} )
 \nonumber \\ && 
 + \frac{a^2}{4} c_2  D_{GW} O\, D_{GW}
 \label{GWimpop2}\nonumber  \\
 C^\calO &\equiv &
 \frac{1}{a^4} \left\langle O M^{-1}\right\rangle +
 \frac{1}{a^4} \left\langle M^{-1} O \right\rangle . 
 \end{eqnarray} 
 $C^\calO$ involves just a single propagator, so it
 is a contact term which only contributes when 
 the operator overlaps with the source or sink, and 
 $ \langle O \rangle/a^4$, with no propagators, is
 a ``double contact term" occurring
 when source, sink and operator all overlap. 

 Two improvement coefficients are free, 
 for example $c_1$ and $c_2$, the others are then determined: 
 \begin{eqnarray}
 c_0 &=& \frac{ 1/2 - c_1 }{1 - a m_0 /2 }
 - \frac{ a m_0 c_2 } {4 (1-a m_0/2)^2 } \nonumber \\
 \lambda_\calO &=& \frac{1 - c_1}{1 - a m_0 /2 }
 - \frac{ a m_0 c_2 } {2 (1-a m_0/2)^2 } \nonumber \\
 \eta_\calO &=& \frac{1 -c_2 - a m_0 /2}{(1 - a m_0 /2)^2 } \, .
  \end{eqnarray}

 The improvement formula is illustrated in Fig.~\ref{fig3pt}. 
 Again we see that the unimproved Ginsparg-Wilson Green's function 
 $\langle M^{-1} O M^{-1} \rangle / a^4$ is no better than the 
 Wilson action Green's function.  It is only by using the 
 improved operator that we can remove all $O(a)$ 
 discretisation errors. 
 
 \section{Conclusions } 

 For Ginsparg-Wilson fermions we can construct
 off-shell improved Green's functions by adding irrelevant operators, 
 and we can write down the improvement coefficients
 in closed form. 

  The nature of the bosonic sector makes no difference, the results are the
 same for Abelian and non-Abelian gauge theories.

 \end{document}